# Starbursts in barred spiral galaxies.

## I. Mrk 712, a new Wolf-Rayet galaxy*

T. Contini[1], E. Davoust[1], S. Considère[2]

[1] URA 285, Observatoire Midi-Pyrénées, 14 avenue E. Belin, F-31400 Toulouse, France
[2] Observatoire de Besançon, B.P. 1615, F-25010 Besançon cedex, France

**Abstract.** We report the discovery of emission from Wolf-Rayet stars in a giant H II region 4.5 arcsec South of the nucleus of the IRAS barred spiral galaxy Mrk 712 (= UGC 5342). The ratio of WNL to OV stars, estimated from the luminosity of the He II$\lambda$4686 line, is 0.2. By comparison with starburst and stellar evolution models, we find that this high value is only compatible with a very young starburst episode (3 − 4 Myr) and a flat initial mass function ($\Gamma = -1$). The presence of the [Ar V] line reveals that the H II region is strongly ionized by the hot Wolf-Rayet stars. The comparison with other barred Wolf-Rayet galaxies suggests that the detection of Wolf-Rayet stars depends on the dust content and orientation of the galaxy.

**Key words:** galaxies: evolution – individual: Mrk 712 – starburst – stellar content – H II regions – stars: Wolf-Rayet

## 1. Introduction

Wolf-Rayet galaxies appear to be the prototype of starburst galaxies where the starburst is extremely young and violent (Conti 1991, Vacca & Conti 1992, Maeder & Conti 1994). The spectral signature of these galaxies is a wide emission line of He II at 4686 Å, attributed to the presence of Wolf-Rayet stars (Allen et al. 1976, Osterbrock & Cohen 1982, Conti 1991). A considerable number of such stars, between $10^2$ and $10^4$, is required to account for this spectrum (Kunth & Sargent 1981, Kunth & Schild 1986, Vacca & Conti 1992). In contrast, there are only about 175 known Wolf-Rayet stars in the whole Galaxy, and 110 in the Large Magellanic Cloud. The ratio of Wolf-Rayet to O stars, measured by Vacca & Conti (1992) for a dozen galaxies, is very high and can reach 0.6; it is compatible with recent models of stellar evolution computed by Maeder & Meynet (1994), provided that star formation occurs in burst mode on very short time scales, less than $10^6$ years.

Wolf-Rayet galaxies might not be very rare, but extremely few of them are known; Conti (1991) gives a list of 40 such galaxies. They have generally been discovered serendipitously, and most of them are Markarian or Zwicky compact galaxies; many also present a disturbed morphology that could be the result of recent gravitational interactions or mergers. They do not stand out by their fundamental properties (morphological type, total luminosity, etc.), except for a very blue continuum and intense emission lines, due to a large number of young, hot and massive OB stars. One reason why so few Wolf-Rayet galaxies are known is probably that the He II line is rather faint compared to the neighboring H$\beta$ and [O III] lines, and spectra with much better signal-to-noise ratio than is usually reached are required for its detection.

It is now essential to enlarge the sample of known Wolf-Rayet galaxies. The detailed study of this type of galaxies is of considerable importance, not only for models of stellar evolution, but also for understanding the starburst phenomenon. Indeed, because Wolf-Rayet stars are the direct offspring of the most massive O stars, and because their lifetime is at most $10^6$ years, the presence of a very large number of such stars in a galaxy provides important constraints on several parameters characterizing starbursts, such as duration and intensity of the burst, time elapsed since the last starburst, initial mass function and upper mass limit of formed stars.

There is also the question of the Wolf-Rayet galaxies as a separate class of galaxies. Do they correspond to a specific stage of the starburst phenomenon, or are the Wolf-Rayet stars only conspicuous when there is no absorption along the line of sight and/or when the orientation of the galaxy is favorable?

## 2. Observations and data reduction

### 2.1. The galaxy Mrk 712

Mrk 712 is a highly inclined barred spiral galaxy. Table 1 gives its catalogue elements, extracted from the Lyon - Meudon Extragalactic Database (LEDA), except for the apparent magnitude (from Mazzarella & Boroson 1993). It has been detected by IRAS, and is slightly less luminous in the far infrared than in the blue. Its only neighbor in a field of 15 arcmin radius is UGC 5344, a Sc galaxy, one magnitude fainter, and 6.2 arcmin away to the South-East, which has a radial velocity of 4106 km

---

, and is thus at about the same distance, but far enough that there is no confusion in the beam of the Nançay antenna.

The galaxy Mrk 712 was included in a survey of IRAS-bright barred Mrk galaxies, a sample of 140 galaxies for which we obtained CCD images, high- and low-resolution spectra and neutral hydrogen profiles. The purpose of the survey was to study the interplay between starbursts and the morphology of galaxies; the results will be published in subsequent papers of this series. After discovering the Wolf-Rayet nature of Mrk 712, we realized that our sample was ideally suited for selecting such galaxies, and carefully inspected all the other spectra obtained in the survey. Five other IRAS barred Mrk galaxies were found to be probable Wolf-Rayet galaxies, although deeper spectroscopy is needed to confirm the classification.

**Table 1.** Catalogue elements of Mrk 712

| | |
|---|---|
| Name | Mrk 712 = UGC 5342 |
| Morphological type | SBbc |
| Right ascension (1950) | $9^h\ 53^m\ 59.1^s$ |
| Declination (1950) | $+15°\ 52'\ 34"$ |
| Apparent blue magnitude | 14.69 |
| Absolute blue magnitude | $-19.72$ |
| Optical radial velocity | $4550 \pm 30$ km s$^{-1}$ |
| Distance (H$_0$= 75 km s$^{-1}$ Mpc$^{-1}$) | 61 Mpc |
| Far infrared luminosity | $1.1 \times 10^{10} L_\odot$ |
| Inclination | $62.9°$ |
| Diameter | 66 arcsec = 19.15 Kpc |

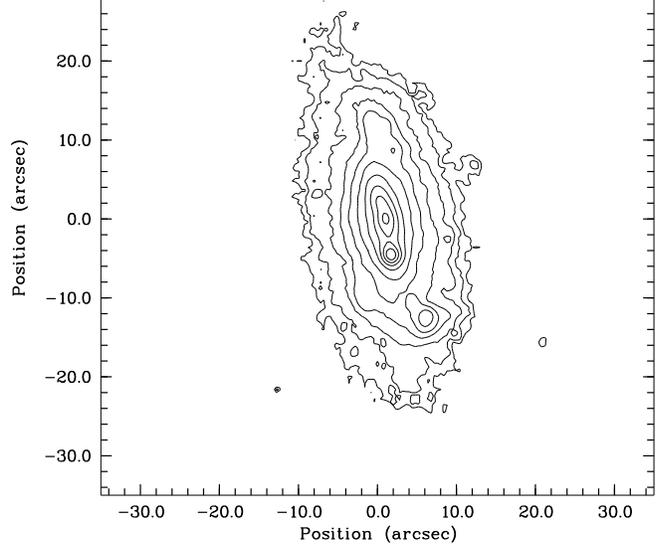

**Fig. 1.** Isophotal contour map of the red CCD image of Mrk 712. The isophotes are separated by 0.5 mag, and the faintest contour is at 23 mag arcsec$^{-2}$. North is up and East is to the left

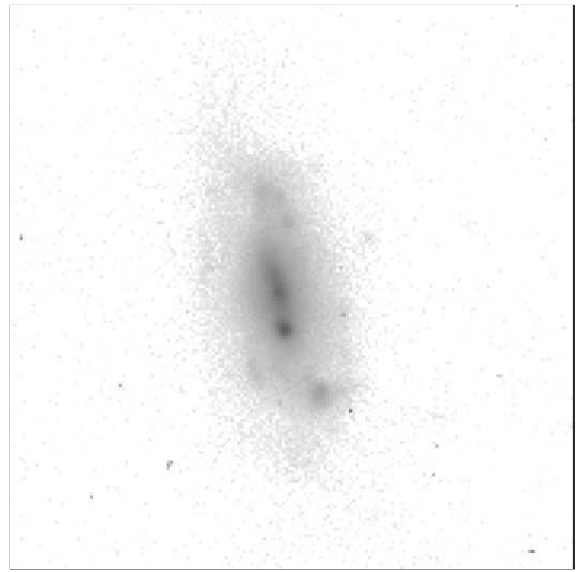

**Fig. 2.** Greyscale map of the red CCD image of Mrk 712. The giant H II region, site of the stellar Wolf-Rayet emission, is located 4.5 arcsec South of the nucleus

### 2.2. CCD imaging

A CCD image was obtained at the 2m telescope (Télescope Bernard Lyot) at Observatoire du Pic du Midi in January 1992. The data were acquired with a 1000$^2$ pixel Thomson CCD (pixel size 0.40 arcsec) and a red (Cousins, $\lambda = 6650$ Å, $\Delta\lambda \simeq 2000$ Å) filter. The exposure time was 20 min. and the seeing (FWHM) 1.3 arcsec. The isophote map is shown on Fig. 1. The condensation 4.5 arcsec South of the nucleus is the seat of the Wolf-Rayet emission. This is the third example of an extranuclear H II region within a barred galaxy where Wolf-Rayet stars have been detected. The other cases are NGC 5430, where Keel (1982, 1987) found WN stars in a bright knot South-East of the center, and the Seyfert galaxy NGC 1365, where Phillips & Conti (1992) detected WC9 stars in a giant H II region located at the end of the bar.

We measured the apparent magnitude of the giant H II region through a 5 arcsec$^2$ aperture. The contribution of the underlying galaxy was estimated and subtracted by measuring the intensity of the symmetrical region with respect to the nucleus with the same aperture. We find an apparent red (Cousins) magnitude of 17.2 mag, which corresponds to a dereddened absolute magnitude of $-17.0$ mag (see below for the extinction value). The giant H II region is unresolved (FWHM = 1.3 arcsec), and we can only give an upper limit of 400 pc to its size ; it is thus neither unusually bright nor unusually large.

### 2.3. Neutral hydrogen line

The neutral hydrogen profile of the galaxy was obtained in October 1993 at the Nançay radiotelescope. At 21cm, the HPBW is 4'(E − W) × 2' (N − S) at declination $\delta = 0°$. The dual channel receiver has a total system temperature of $\simeq 40$ K. The 1024-channel autocorrelation spectrometer, covering a total bandwidth of 6.4 MHz, was split into two banks for the two orthogonal linear polarizations, providing a channel spacing of 2.6 km s$^{-1}$. The spectrum was obtained by 2 min. scans with spatial on-off switching, and was smoothed to a resolution of 8 km s$^{-1}$, after combining the two polarizations.

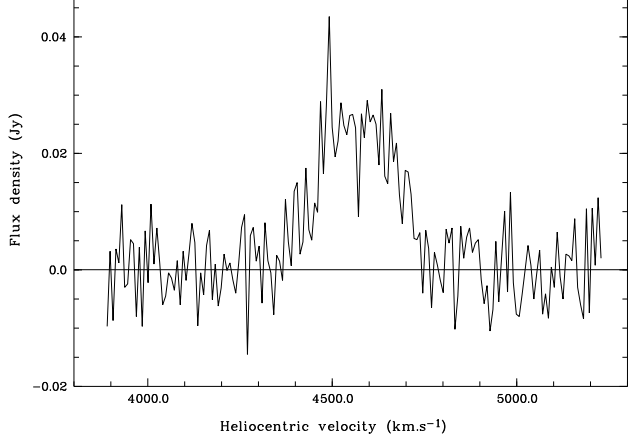

**Fig. 3.** H I profile of Mrk 712

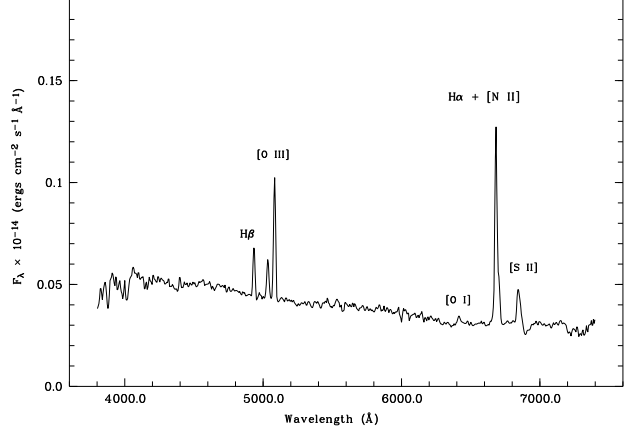

**Fig. 4.** Dereddened optical spectrum of the nucleus of Mrk 712

The detection was obtained with a typical rms noise of 5 mJy at the resolution of 8 km s$^{-1}$. This corresponds to a total integration time of about 75 min. The H I profile of Mrk 712 is shown in Fig. 3.

The measured heliocentric radial velocity of the galaxy is 4570 km s$^{-1}$, in good agreement with the optical radial velocity (Table 1). The line widths at 20% and 50% of maximum intensity are equal to 295 km s$^{-1}$ and 200 km s$^{-1}$ respectively. The total mass of the galaxy, derived from the width at 20% and the diameter of the galaxy, is $5 \times 10^{10} M_\odot$, and corresponds to a mass to blue light ratio of 5. The total flux of the H I line is 6.4 Jy km s$^{-1}$, and the total H I mass $5.6 \times 10^9$ M$_\odot$. This corresponds to a ratio of H I mass to blue luminosity of 0.49, equal to the mean value found for FIR-bright Sbc galaxies (Fournou & Davoust 1995). The "bell-shaped" H I profile of Mrk 712 stands out from the approximately symmetrical double-peaked profile of the other known barred Wolf-Rayet galaxies, but perhaps this is due to the mediocre S/N of our data.

### 2.4. Long-slit spectroscopy

A low-resolution spectrum of Mrk 712 was obtained on the night of 1994 January 13 - 14 at the 193cm telescope of Observatoire de Haute-Provence. The data were acquired with the CARELEC spectrograph (Lemaître et al. 1990) and a $512^2$ pixel Tektronix CCD. The spectral resolution was 260Å/mm, and provided a spectral coverage of 3800 Å – 7400 Å with a resolution $\Delta\lambda \sim 15$ Å as measured from the FWHM of gaussian profiles adjusted to the He comparison lines. During the same night we also observed the spectrophotometric standard star Hiltner 600 taken from the list given by Massey et al. (1988) in order to flux calibrate the spectrum of the galaxy. Spectra of He comparison lines were obtained immediately before and after the galaxy integration in order to calibrate accurately the wavelength scale. The slit (width 2.7 arcsec) was positioned along the major axis of the galaxy (PA = 8°) and the exposure time was 2700 sec. The seeing (FWHM) was 2.6 arcsec.

The spectroscopic data were reduced according to a standard reduction procedure using the MIDAS package LONG. These included bias subtraction, flat-field corrections, wavelength calibration, and sky subtraction.

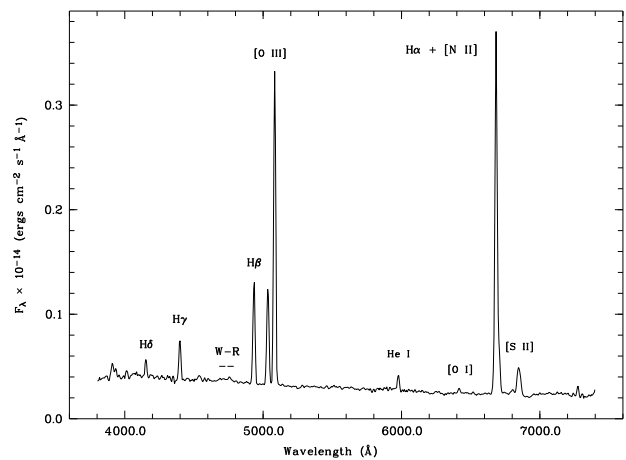

**Fig. 5.** Dereddened optical spectrum of the giant H II region of Mrk 712. The spectral signature of Wolf-Rayet stars (W-R) is very weak compared to the intense emission lines of the ionized gas. Note the difference in flux scale with fig. 4

The two-dimensional spectrum of Mrk 712 was then corrected for foreground extinction in our own Galaxy. We accounted for the foreground reddening using the value of the total Galactic extinction E(B-V) = 0.012 given by Burstein & Heiles (1984) and the average Galactic extinction law of Howarth (1983) with $R_V = 3.1$, covering entirely the spectral range of interest.

We extracted one-dimensional elementary spectra by adding several columns (spatial dimension). One elementary spectrum corresponds to the nucleus of the galaxy (addition of 3 columns, or $\simeq 3.4$ arcsec), the other one to the extranuclear emission-line region 4.5 arcsec away from the nucleus along the bar, also averaged over 3 columns, since the He II line was clearly detected on 3 columns. All data referring to the first elementary spectrum are designated by the subscript $N$, and those referring to the giant H II region by the subscript $HII$. The two elementary spectra were cleaned of radiation events ("cosmic rays") before further analysis.

Because of the low spectral resolution, the H$\alpha$ + [N II] complex and the [S II]$\lambda\lambda 6716, 6731$ lines are blended. We fit multi-gaussian profiles to the observed profile in order to measure accurately the individual emission lines. In the procedure for deblending H$\alpha$ + [N II] we only set the relative position of the lines whereas, for deblending the weaker [S II] complex, we furthermore constrain the width of the two nebular lines to be equal. We did not want to make any assumptions about the physical conditions in the ionized gas ($n_e$ and $T_e$); the line intensities were thus considered as free parameters. Line fluxes and equivalent widths were determined using MIDAS standard commands. The fluxes of all emission lines were measured by direct integration under the line profile. The relative uncertainty is about 10% for the most intense lines (H$\alpha$ and [O III]), and about 20% for the others.

## 3. Analysis of the spectroscopic data

### 3.1. Absorption, reddening and line fluxes

All observed line fluxes were first corrected for the color excess due to internal extinction. The effect of reddening is commonly parameterized as

$$\frac{I(\lambda_1)}{I(\lambda_2)} = \frac{F(\lambda_1)}{F(\lambda_2)} 10^{c[f(\lambda_1)-f(\lambda_2)]} \tag{1}$$

where c is a reddening constant proportional to $E_{B-V}$, $f(\lambda)$ is the standard Whitford (1958) reddening curve parameterized by Miller & Mathews (1972), $F(\lambda)$ is the observed flux, and $I(\lambda)$ is the intrinsic, unreddened flux. Although several lines of the Balmer series are present in the spectrum of the giant H II region of Mrk 712, we chose to use only the H$\alpha$ and H$\beta$ lines to determine the internal extinction. This decision was based on the fact that the spectrum of Mrk 712 exhibits broad absorption wings centered on the narrower, nebular Balmer emission lines. We interpret these absorption wings as underlying stellar Balmer absorption. This implies that the observed fluxes of the nebular Balmer emission lines may be underestimated, and the internal extinction overestimated. It is reasonable to adopt a constant value for the equivalent width of the Balmer absorption lines. McCall et al. (1985) measured an equivalent width due to Balmer absorption equal to 1.8 Å for H II regions in nearby galaxies. Keel (1983) found that the equivalent width due to stellar absorption at H$\alpha$ and H$\beta$ ranges from 1 to 3 Å in the nuclei of bright spiral galaxies. We adopt a value $W_\lambda^{abs} \simeq 2$ Å. The Balmer emission-line equivalent widths were corrected by $W_\lambda^{corr} = W_\lambda^{obs} - W_\lambda^{abs}$, and since, by definition, $F_\lambda = -W_\lambda \times f_\lambda$, where $f_\lambda$ is the flux in the continuum reduced to a width of 1 Å, the appropriate correction to the Balmer line fluxes is

$$F_\lambda^{corr} = F_\lambda^{obs}\left(1 - \frac{W_\lambda^{abs}}{W_\lambda^{obs}}\right) \tag{2}$$

The reddening parameter $c_{H\beta}$ was estimated using the observed flux ratio $F(H\alpha)/F(H\beta)$ corrected for Balmer absorption and by assuming that the intrinsic flux ratio $I(H\alpha)/I(H\beta)$ is known. We derive the electron densities $n_e$ from the [S II] $\lambda6716/\lambda6731$ flux ratio. We find an electron density, $n_e$, equal to 130 cm$^{-3}$ for both the giant H II region and the nucleus of Mrk 712. The error on this parameter follows from the uncertainty in the relative line fluxes. A 20% uncertainty in the [S II]$\lambda6716/\lambda6731$ flux ratio corresponds to an uncertainty in the determination of $n_e$ of about 100 cm$^{-3}$. The electron density in the two emission-line regions is typically that of an H II region. However, we could not derive the electron temperature $T_e$ from the [O III]$\lambda4959+\lambda5007/\lambda4363$ flux ratio because the [O III]$\lambda4363$ line does not appear in the spectra. Therefore, the case B Balmer decrement was adopted in H II emission regions, $I(H\alpha)/I(H\beta) = 2.85$, assuming an electronic temperature $T_e = 10^4$ K (Brocklehurst 1971, Osterbrock 1989). The relation used to calculate the reddening parameter $c_{H\beta}$ is

$$c_{H\beta} = \frac{\log[I(H\alpha)/I(H\beta)] - \log[F(H\alpha)/F(H\beta)]}{f(H\alpha) - f(H\beta)} \tag{3}$$

where $f(\lambda) = \delta m_\lambda/\delta m_\beta$ is the value of extinction at wavelength $\lambda$ relative to that at H$\beta$ (Miller & Mathews 1972). We find an internal extinction $c_{H\beta}$ of 0.28, close to the value of 0.26 given by LEDA. The final intrinsic fluxes are then given by the relation

$$I(\lambda) = F(\lambda) 10^{c_{H\beta} f(\lambda)} \tag{4}$$

Values of the observed flux $F(\lambda)$ (corrected for Balmer absorption) and intrinsic flux $I(\lambda)$ are listed in Tables 2 and 3 for the two elementary spectra.

**Table 2.** Intensity of nebular emission lines in the nucleus and the giant H II region of Mrk 712

| Line | $\lambda$ (Å) | $\delta m_\lambda$[1] (mag) | Observed[2] $F_N$ | $F_{H\text{II}}$ | Corrected[2] $I_N$ | $I_{H\text{II}}$ | Relative[3] $R_N$ | $R_{H\text{II}}$ |
|---|---|---|---|---|---|---|---|---|
| H$\delta$ | 4102 | 1.418 | ... | 0.30 | ... | 0.66 | ... | 18.36 |
| H$\gamma$ | 4340 | 1.361 | ... | 0.76 | ... | 1.59 | ... | 44.15 |
| He I | 4471 | 1.315 | ... | 0.07 | ... | 0.15 | ... | 4.15 |
| H$\beta$ | 4861 | 1.182 | 0.52 | 1.89 | 0.97 | 3.60 | 100.00 | 100.00 |
| [O III] | 4959 | 1.152 | 0.34 | 1.84 | 0.62 | 3.45 | 64.51 | 95.78 |
| [O III] | 5007 | 1.138 | 1.04 | 5.48 | 1.89 | 10.19 | 195.85 | 283.04 |
| He I | 5876 | 0.919 | ... | 0.24 | ... | 0.40 | ... | 11.00 |
| [O I] | 6300 | 0.835 | 0.08 | 0.08 | 0.12 | 0.13 | 12.84 | 3.50 |
| [N II] | 6548 | 0.790 | 0.04 | 0.10 | 0.06 | 0.15 | 6.27 | 4.27 |
| H$\alpha$ | 6563 | 0.788 | 1.83 | 6.72 | 2.77 | 10.33 | 287.10 | 286.80 |
| [N II] | 6583 | 0.784 | 0.35 | 0.71 | 0.53 | 1.09 | 54.72 | 30.24 |
| [S II] | 6717 | 0.762 | 0.25 | 0.41 | 0.37 | 0.63 | 38.63 | 17.46 |
| [S II] | 6731 | 0.759 | 0.19 | 0.31 | 0.28 | 0.47 | 29.32 | 13.07 |

[1] Extinction at wavelength $\lambda$
[2] Intensity in units of $10^{-14}$ erg cm$^{-2}$ s$^{-1}$
[3] Corrected for extinction and normalized to $I(H\beta) = 100$

### 3.2. Velocity profile of Mrk 712

The emission lines can be seen over a large range along the bar of the galaxy (over 34 arcsec for H$\alpha$), enabling us to derive a velocity profile from the position of up to ten emission lines with S/N ratio greater than 5 and weights according to their intensity (see Fig. 6). The very small velocity gradient along the bar (which is aligned with the major axis) means that the gas flows along the bar rather than around the galactic center, except at the location of the giant H II region. The latter strongly perturbs the galactic dynamics, as it moves away from us with a velocity of 80 km s$^{-1}$ with respect to the bar. The velocity gradient is negative from North to South, and increases notably beyond the H II region; this means that the Eastern side of the galaxy is the near side, if the spiral arms are trailing.

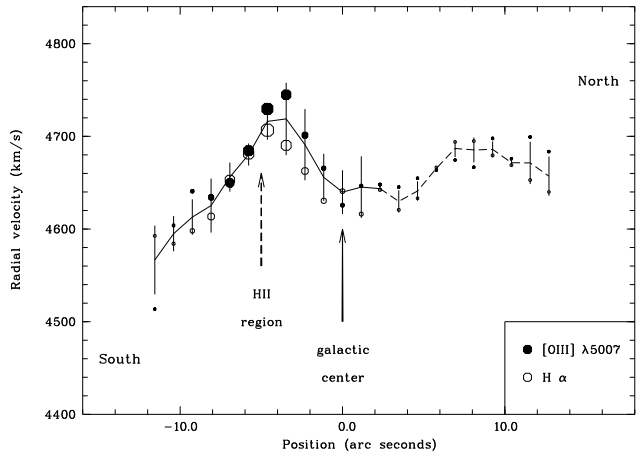

**Fig. 6.** Heliocentric velocity profile of Mrk 712, from 4 to 10 emission lines (solid curve) and from the two brightest emission lines (dashed curve). The velocities of H$\alpha$ and [O III] are plotted with symbol sizes proportional to the emission-line intensity.

### 3.3. Oxygen abundance and metallicity

Oxygen is an important element for our subsequent analysis, as it is used to define the metallicity of each region. Since the [O II]$\lambda$3727 line is not present in our elementary spectra, the oxygen abundance was determined from the empirical relationship between log(O/H) and the intensities of the nebular [O III] lines given by Edmunds & Pagel (1984) with the following linear fit given by Vacca & Conti (1992)

$$\log (O/H) = -0.69 \log R_3 - 3.24 \quad (-0.6 \leq \log R_3 \leq 1.0), \quad (5)$$

where

$$R_3 \equiv \frac{I([O\,III]\lambda 4959) + I([O\,III]\lambda 5007)}{I(H\beta)} \quad (6)$$

This equation refers to the so-called upper-branch of the empirical oxygen abundance calibration. The oxygen abundance in the H II region and the nucleus of Mrk 712 are found to be respectively equal to 0.27 and 0.36 $Z_\odot$, thus rather metal poor. For the solar oxygen abundance we adopted the local galactic value of log(O/H)$_\odot = -3.08$ given by Meyer (1985). Edmunds & Pagel (1984) estimate the intrinsic uncertainty in the calibration to be about $\pm 0.2$ in log(O/H); in what follows, we adopt a mean value of (O/H) = 0.3 $Z_\odot$ for the entire galaxy.

### 3.4. Emission lines in the Wolf-Rayet "bump"

In the giant H II region of Mrk 712, the spectrum shows the set of emission lines between 4650 and 4850 Å often called the Wolf-Rayet "bump" (Kunth & Schild 1986, Conti 1991). In all three columns of the H II region spectrum we see two broad emission lines (FWHM $\sim$ 32 Å), one is identified with He II$\lambda$4686, the other is uncertain, corresponding to either N III$\lambda$4640 or C III/IV$\lambda\lambda$4650,4658. The difficulty of identifying this second broad emission line is mainly due to our low spectral resolution.

In addition to these broad emission features, we detect with a high confidence level three narrower (FWHM $\sim$ 18 Å) nebular lines, namely [Ar IV]$\lambda\lambda$4711,4740 and [Ar V]$\lambda$4625. Figure 7 shows the integrated spectrum of the giant H II region in the range of the Wolf-Rayet "bump". The nebular He I$\lambda$4471 emission line appears as broad as the He II$\lambda$4686 line on Fig. 7 because it is blended with a cosmic ray which increases its line width and redshifts its position. After decomposition, the He I$\lambda$4471 line has a width of about 18 Å, comparable to that of the other nebular lines, and is at the expected position.

Because the ionization potential of Ar V is very high (75 eV), the presence of the [Ar IV]$\lambda\lambda$4711,4740 and [Ar V]$\lambda$4625 emission lines requires a very hard ionizing spectrum. In these conditions, a nebular contribution to the He II$\lambda$4686 emission line is expected, since the ionization potential of He II is 54.4 eV. This is supported by Fig. 7 where the He II$\lambda$4686 feature appears to be composed of a narrow component atop a broader base. We measured the intensity of these different lines by multi-gaussian fitting, since the lines are not quite resolved. The relative uncertainty is about 30% for the most intense line (He II$\lambda$4686), and about 50% for the others. The intensities before and after extinction correction are given in Table 3.

**Table 3.** Intensity of emission lines in the Wolf-Rayet "bump" of the giant H II region of Mrk 712

| Line | $\lambda$ (Å) | $\delta m_\lambda$ (mag) | Observed[1] $F_{H\,II}$ | Corrected[1] $I_{H\,II}$ | Relative[2] $R_{H\,II}$ |
|---|---|---|---|---|---|
| [Ar V][4] | 4625 | 1.260 | 0.03 | 0.07 | 1.82 |
| C III/IV + N III[3] | 4650 4640 | 1.252 | 0.08 | 0.17 | 4.62 |
| He II[3] | 4686 | 1.239 | 0.09 | 0.18 | 4.86 |
| He II[4] | 4686 | 1.239 | 0.03 | 0.06 | 1.80 |
| [Ar IV][4] | 4711 | 1.231 | 0.02 | 0.03 | 0.92 |
| [Ar IV][4] | 4740 | 1.221 | 0.03 | 0.05 | 1.35 |

[1] Intensity in units of $10^{-14}$ erg cm$^{-2}$ s$^{-1}$
[2] Corrected for extinction and normalized to $I(H\beta) = 100$
[3] emission from Wolf-Rayet stars (FWHM = 32 Å)
[4] nebular emission (FWHM = 18 Å)

### 3.5. Source of the hard ionizing spectrum

The two elementary spectra of Mrk 712 exhibit "blue" continua and strong nebular forbidden and recombination lines (see Figs. 4 and 5). As a check of the source of ionization in these regions, we calculated the reddening-corrected line intensity ratios I([O I]$\lambda$6300)/I(H$\alpha$), I([N II]$\lambda$6584)/I(H$\alpha$), and I([S II]$\lambda$6716 + $\lambda$6731)/I(H$\alpha$). When plotting these values against the corresponding value of I([O III]$\lambda$5007)/I(H$\beta$) we find that the intensity ratios determined from our elementary spectra are typical of those found for H II regions and starburst galaxies (Veilleux & Osterbrock 1987) and are at ionization levels comparable to those of other Wolf-Rayet galaxies (Vacca & Conti 1992). Hot stars are indeed the ionizing sources in the two emission-line regions of Mrk 712. But the nucleus is closer to the narrow-line active galactic nuclei region than the giant H II region; this implies slightly different ionization conditions in these two regions.

Nevertheless, the giant H II region is rather unusual, because a very energetic source of photons ($\lambda \leq 228$ Å) is needed for indirectly exciting the [Ar IV]$\lambda\lambda$4711,4740 and [Ar V]$\lambda$4625 collisional emission lines, and *a fortiori* the nebular He II$\lambda$4686 recombination line. Extragalactic H II regions presenting such a high degree of ionization have already been observed in H II galaxies (e.g. Bergeron 1977, Campbell et al. 1986, Kunth & Schild 1986, Vacca & Conti 1992, and Motch et al. 1993). It is not a coincidence that the broad He II$\lambda$4686 line characteristic of Wolf-Rayet stars is also present in the spectra of these galaxies.

Assuming that the ionizing source in the giant H II region of Mrk 712 is a cluster of hot stars emitting EUV photons, rather than binary stars emitting X-rays (Motch et al. 1993), the question is : what kind of stars ?

The effective temperature $T_{eff}$ of the ionizing source can be derived from the ratio $Q(He^+)/Q(H^0)$ between the $He^+$ and $H^0$ ionizing photons. Diagrams giving $T_{eff}$ vs $Q(He^+)/Q(H^0)$ for realistic model stellar atmospheres can be found in Garnett et al. (1991) and Gabler et al. (1992). From the fluxes in the nebular lines of He II$\lambda$4686 and H$\beta$, we find a ratio $Q(He^+)/Q(H^0) = 10^{-2}$. This gives an effective temperature between 70 000 and 90 000 K, which corresponds to evolved Of stars or, more likely, to Wolf-Rayet stars at an advanced stage, WNE, WC or WO (Maeder & Conti 1994).

This is corroborated by observations of gaseous nebulae around Wolf-Rayet stars, which have revealed the existence of an extended strongly ionized region ($He^{2+}$), for example around a WO3 star in IC 1613 (Davidson & Kinman 1982, Garnett et al. 1991, Kingsburgh & Barlow 1995), around a WN1 star in LMC N79W (Pakull 1991), or two stars WN1 + O6 IIIf in SMC N76 (Garnett et al. 1991). Such regions are believed to have been strongly heated and ionized by Wolf-Rayet stars whose effective temperature is sufficiently high to produce a flux of energetic $He^+$ Lyman continuum photons.

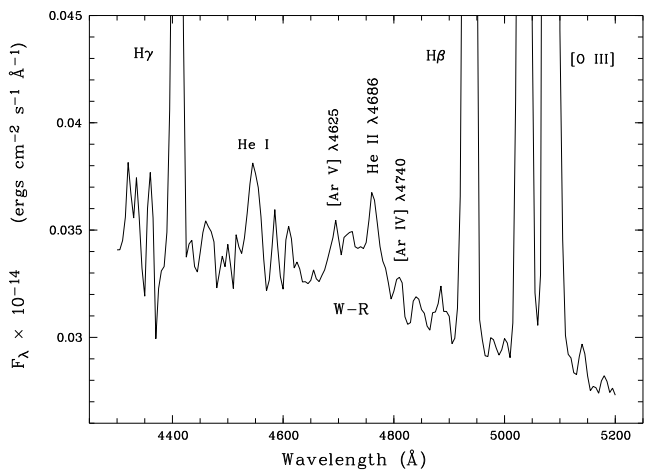

**Fig. 7.** The Wolf-Rayet "bump" in the dereddened optical spectrum of the giant H II region of Mrk 712. The He II$\lambda$4686 emission line consists of a narrow component due to nebular emission atop a broader component due to emission from Wolf-Rayet stars. The [Ar IV]$\lambda\lambda$4711,4740 and [Ar V]$\lambda$4625 emission lines indicate a very hard ionizing spectrum.

*3.6. Population of hot and massive stars*

Two broad (FWHM $\simeq$ 32 Å) emission lines are present in the spectrum of the giant H II region of Mrk 712. One is the He II$\lambda$4686 line, the other is probably a blend of several lines like N III$\lambda$4640 and C III/IV$\lambda\lambda$4650,4658. The same type of spectral signature is present in the forty or so known Wolf-Rayet galaxies (Conti 1991) and attributed to the presence of a large number of Wolf-Rayet stars. Mrk 712 can thus be classified as a Wolf-Rayet galaxy.

A few hot WC stars are probably necessary for producing the high degree of ionization observed in the giant H II region. But the absence of the C IV$\lambda$5808 emission line indicates that they are not in sufficient number to be detected in our spectrum. The strength of the He II$\lambda$4686 emission line indicates that most Wolf-Rayet stars present are of type WN (Vacca & Conti 1992). The identification of the subgroup of WN stars in Wolf-Rayet galaxies is much more difficult than that of WC stars, because of the intrinsic faintness of the nitrogen lines (N III$\lambda$4640, N IV$\lambda$4057 and N V$\lambda$4603) compared to that of He II$\lambda$4686 (Conti & Massey 1989). The resolution and signal-to-noise ratio of our spectrum did not allow us to detect these lines with enough certainty to identify the WN subgroup; following other investigators (Osterbrock & Cohen 1982, Armus et al. 1988, Vacca & Conti 1992), we thus assume that most Wolf-Rayet stars in the giant H II region of Mrk 712 are of type WNL.

We now determine the ratio of Wolf-Rayet to O-type stars in the giant H II region; this is one of the most important observational quantities, as it allows a comparison between our observations and models of starburst and stellar evolution. To this end, we use the method developed by Vacca (1991), and summarized by Vacca & Conti (1992).

We first compute the luminosity $L_\lambda$ (in erg s$^{-1}$) of the emission lines, which is simply $4\pi D^2 I(\lambda)$, where $D$ is the distance and $I(\lambda)$ is the line intensity after correction for internal extinction. The results are given in Table 4 for the giant H II region, and can be derived from col. 6 of Table 2 for the nucleus. The H$\alpha$ luminosity of the nucleus is comparable to that of 30 Doradus, whereas the H$\alpha$ luminosity of the giant H II region is three times higher and comparable to that of giant H II regions in M82 (Kennicutt 1984).

We use the luminosity of the He II$\lambda$4686 broad line to estimate the number of WN stars. According to Vacca & Conti (1992), it is preferable to use only one emission line rather than all those of the Wolf-Rayet "bump", from 4600 to 4700 Å (Kunth & Joubert 1985, Mas-Hesse & Kunth 1991). The typical luminosity of the He II$\lambda$4686 line in a single WNL star is provided by Vacca (1991). Using this value and $L$(He II$\lambda$4686) derived above, we estimate that there are about 450 WN stars in the giant extranuclear H II region of Mrk 712, a relatively high amount by comparison with the numbers estimated in the emission-line regions of other Wolf-Rayet galaxies.

Next, using the equations given by Vacca & Conti (1992), we compute the ratio between the number of WN and "reference" O7V stars, $N_{WNL}/N'_{O7V}$, using the observed ratio of $L(H\beta)/L(He II\lambda 4686)$ and assuming that the ionizing photons necessary to produce the observed recombination spectrum are produced by the O-type and Wolf-Rayet stars.

We finally estimate the total number of O-type stars and the ratio $N_{WN}/N_O$ in the giant H II region, in order to compare these values with those predicted by the models of stellar evo-

**Table 4.** Observational quantities used to characterize the starburst in the giant H II region of Mrk 712

| | |
|---|---|
| $L(\text{H}\alpha)^1$ | 45.5 |
| $W_\lambda(\text{H}\alpha)$ (Å) | $-320.0$ |
| $L(\text{H}\beta)^1$ | 15.9 |
| $W_\lambda(\text{H}\beta)$ (Å) | $-63.0$ |
| $L(\text{He II}\lambda 4686)^{1,2}$ | 0.8 |
| $W_\lambda(\text{W-R "bump"})$ (Å) | $-5.4$ |
| $L(\text{W-R "bump"})/L(\text{H}\beta)$ | 0.1 |
| $N_{WNL}$ | 450.0 |
| $N_{WNL}/N'_{O7V}$ | 0.18 |
| $\eta_0$ | $1.0 \pm 0.4$ |
| $N_{OV}$ | $2500 \pm 900$ |
| $N_{WN}/N_{OV}$ | $0.2 \pm 0.1$ |

[1] Luminosities are expressed in units of $10^{39}$ erg s$^{-1}$
[2] Luminosity of the stellar emission line only

lution of Maeder (1991) and Maeder & Meynet (1994). To do this, we calculate the quantity $\eta_0 \equiv N'_{O7V}/N_{OV}$ where $N_{OV}$ is the total number of OV stars present (Vacca 1994). For a given value of $\eta_0$ we can then convert the number of equivalent O7V stars derived from the spectral analysis into the total number of OV stars. The conversion factor $\eta_0$ depends primarily on three parameters: the slope and upper mass limit of the stellar mass function and the metallicity of the H II region (Vacca 1994).

To evaluate the parameter $\eta_0$, we adopted the observed oxygen abundance log $(Z/Z_\odot) = -0.5$ as the "metallicity" and used models of Vacca (1994) for an upper mass limit $M_{upp} = 120~M_\odot$ and different values for the slope of the stellar mass function between $-1.5$ and $-3$ (the slope for a Salpeter mass function is $-2.35$). In their study of a dozen Wolf-Rayet galaxies, Vacca & Conti (1992) arbitrarily chose a slope of $-2.5$, which limits the possible values of $\eta_0$.

The values of $\eta_0$, $N_{OV}$ and $N_{WN}/N_{OV}$ are given in Table 4. The uncertainties on these values are due to the different possible values of the slope of the stellar mass function. We find quite a large ratio of WNL to O stars in the giant H II region of Mrk 712.

### 3.7. Uncertainties

The first source of uncertainty is the size of the giant H II region. The CCD image gives a full width at 10% of 3 arcsec, comparable to the spatial extent of the elementary spectrum. This is an upper limit to the size, since the H II region is unresolved, but there was no point in taking a smaller extraction region for the spectrum, since the seeing was not very good during the spectroscopic observations. The elementary spectrum for the giant H II region might thus include unwanted signal, and the number of O-type stars might thus be overestimated.

The assumption that there are a majority of WNL stars rests on the fact that our spectrum does not allow one to distinguish the different types of WN stars. But there must be WNE or even WC stars in the giant H II region (see Sect. 3.5), which, to some extent, will increase the $N_{WR}/N_O$ ratio.

We estimated the metallicity from the oxygen abundance. But the latter is rather uncertain; since we did not observe the [O II]$\lambda 3727$ line, we had to rely on the calibration by Edmunds & Pagel (1984), which has its own uncertainties.

Finally, Maeder & Meynet (1994) have shown that increased mass loss during stellar evolution will increase the $N_{WNL}/N_{OV}$ ratio when the metallicity is larger than 0.02 $Z_\odot$. This strong dependence of $N_{WNL}/N_{OV}$ on mass loss is still not well quantified, and may explain remaining discrepancies between theoretical predictions and observations.

## 4. Discussion

In this section, we discuss the properties of the giant H II region of Mrk 712 in the context of stellar evolution and starburst models, and how they fit in with observations of other Wolf-Rayet barred galaxies.

### 4.1. Age and Initial Mass Function (IMF) of the starburst

We used recent models of evolutionary population synthesis in starburst regions (Mas-Hesse & Kunth 1991, Cerviño & Mas-Hesse 1994, Leitherer & Heckman 1995) to characterize the star formation episode (age and IMF) in the nucleus and in the giant H II region of Mrk 712. The observational quantities used are summarized in Table 4. The models depend strongly on metallicity, for which the value derived above, $Z = 0.3~Z_\odot$, is adopted.

We only have two observational quantities, $W_\lambda(\text{H}\beta) = -12$ Å and $W_\lambda(\text{H}\alpha) = -58$ Å, to date the starburst in the nucleus of Mrk 712. We find a good agreement between observations and models for a Salpeter IMF ($\Gamma = -2.35$ and $M_{upp} = 100$ - $120~M_\odot$) and an age of the burst between 7 and 9 Myr.

In the giant H II region, the constraints on the age and the IMF of the starburst are more important. In addition to the equivalent widths of the nebular lines, which determine the ionizing flux, there are the Wolf-Rayet stars, which provide a direct constraint on the stellar population through $W_\lambda(\text{W-R "bump"})$ and $L(\text{W-R "bump"})/L(\text{H}\beta)$. The best fit between all observational quantities and the models are found for a flat IMF ($\Gamma = -1$ and $M_{upp} = 100$ - $120~M_\odot$) and for an age of the burst between 3 and 4 Myr. Despite the high uncertainty in the determination of the luminosity of the Wolf-Rayet "bump" (as high as a factor 2), a steeper IMF cannot explain the high number of Wolf-Rayet stars found in this H II region.

This result is corroborated by new evolutionary population synthesis models which have the advantage of including the influence of Wolf-Rayet stars. In Fig. 8 we compare the observed $N_{WNL}/N_{OV}$ ratio and its dependence on oxygen abundance (in solar units) with predictions of stellar evolution models. The observations are from this paper (filled square) and from Vacca & Conti (1992) (open symbols). The three models (lines) shown in Fig. 8 do predict that the $N_{WNL}/N_{OV}$ ratio increases with metallicity, but there are large differences as to the actual values predicted. The first model of continuous star formation (dashed line) developed by Maeder (1991) predicts a ratio that is ten times smaller than observed in Wolf-Rayet galaxies. More recent models by Maeder & Meynet (1994), where star formation occurs in burst mode, are in better agreement with the observations. However, the observations of all the Wolf-Rayet galaxies cannot be reconciled with a unique slope for the IMF. We plot predictions for two values of the IMF slope ($\Gamma = -1$ and $-2$) in Fig. 8. The observed high $N_{WNL}/N_{OV}$ ratio for Mrk 712 is well fitted by the Γ = −1 curve. We also used the most recent evolutionary population synthesis models developed by Meynet (1995) who studies the Wolf-Rayet population in starburst galaxies. The comparison between the observed $L(\text{He\,{\sc ii}}\lambda 4686)/L(\text{H}\beta)$ ratio and that predicted by a stellar model with a high stellar mass loss rate gives an age of the burst between 3 and 3.5 Myr with a flat IMF (Γ = −1).

In conclusion, all evidence points to a flat IMF (Γ = −1) and an age of 3 to 4 Myr for the starburst in the giant H II region of Mrk 712 whereas the star formation episode in the nucleus is two times older with a Salpeter IMF.

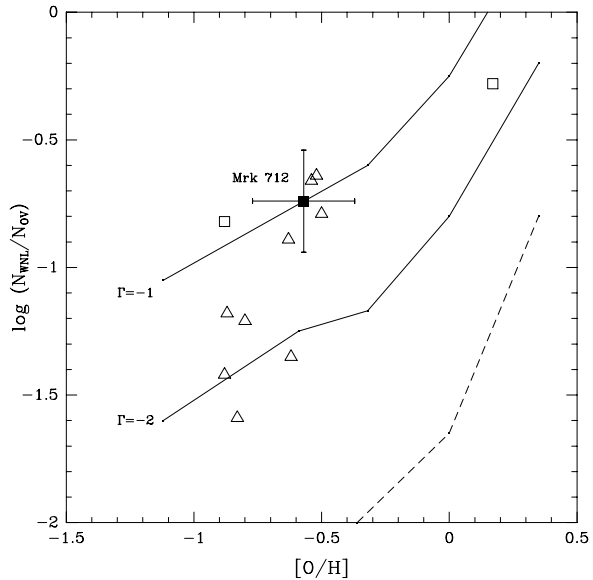

**Fig. 8.** Ratio of WNL to OV stars vs oxygen abundance in solar units. The open symbols denote the Wolf-Rayet galaxies of Vacca & Conti (1992). Squares are barred spiral galaxies. The filled square is the giant H II region of Mrk 712 with its uncertainty. The dashed line denotes the predictions for continuous star formation; the solid lines are for values estimated for a burst. Γ is the slope of the initial mass function. The models of stellar evolution are from Maeder & Meynet (1994)

### 4.2. The class of Wolf-Rayet barred spiral galaxies

Seven of the currently known Wolf-Rayet galaxies are barred spirals. They are summarized in Table 5, which gives their morphological type, inclination, distance, and the ratios of far infrared luminosity and neutral hydrogen mass to blue luminosity. Mrk 712 is by far the most distant galaxy of the sample.

Most galaxies have been detected by IRAS, and have a far infrared luminosity comparable (within a factor 2) to the total blue luminosity. The mean value of the ratio of the two luminosities is 0.7, significantly smaller than the value of 2.45 found by Deutsch & Willner (1987) for a sample of 102 starburst galaxies. This is an indication that these galaxies might contain less dust than other starburst galaxies, and could explain why Wolf-Rayet stars were detected. However the sample is too small for definite conclusions.

The ratio of hydrogen mass to blue luminosity is lower than the average for the corresponding morphological type (Fournou & Davoust 1995) by a factor 2 (− 0.32 in logarithmic units) ; Mrk 712 is the only galaxy for which this ratio is normal.

**Table 5.** Characteristics of the class of Wolf-Rayet barred spiral galaxies

| Name | type | i (°) | D (Mpc) | $\log(\frac{L_{fir}}{L_{blue}})$ | $\log(\frac{M_{HI}}{L_{blue}})$ |
|---|---|---|---|---|---|
| NGC 1365 | SBbc | 57.1 | 19 | 0.03 | −0.62 |
| Mrk 710 | SBb | 50.6 | 20 | −0.38 | −0.45 |
| Mrk 712 | SBbc | 62.9 | 61 | −0.37 | −0.31 |
| Mrk 52 | SBa | 58.2 | 28 | −0.10 | −0.95 |
| Pox 139 | SBc | 48.8 | 27 | | −0.65 |
| Mrk 799 | SBb | 51.7 | 43 | 0.14 | −0.87 |
| NGC 6764 | SBbc | 58.1 | 36 | −0.27 | −0.86 |
| mean (σ) | | 55.3 | | −0.16 | (−0.32) |

Surprisingly, all the galaxies are highly inclined (between 48.8° and 62.9°); this trend is also present among our Wolf-Rayet galaxy candidates and suggests that orientation plays a role in the detectability of such galaxies. But the only galactic-scale effect that we can think of, supernova driven winds, should be a more efficient vacuum-cleaner perpendicular to the disk of the galaxy, and should favor the detectability of face-on Wolf-Rayet galaxies.

We thus have another reason to pursue the search for other such galaxies, to understand whether the trends uncovered here, low far infrared luminosity, low relative hydrogen content, high inclination, are really important or not.

*Acknowledgements.* We are grateful to Daniel Schaerer for stimulating discussion and to the referee, W.D. Vacca, for helpful and constructive comments. We thank the TAC of Observatoire de Nançay for generous time allocation and for helpful advice. We also thank the staff of Observatoire du Pic du Midi, Observatoire de Haute-Provence and Observatoire de Nançay for assistance at the telescope.